\documentclass[reprint,prX,amsmath,amssymb]{revtex4-1}

\usepackage{graphicx}
\pdfoutput=1
\usepackage{dcolumn}
\usepackage{bm}
\usepackage{multirow}
\usepackage{float}
\usepackage{wrapfig}
\usepackage[normalem]{ulem}
\usepackage{color}
\usepackage{xr}
\usepackage{booktabs}
\usepackage{longtable}
\usepackage[colorlinks=true, citecolor=blue, urlcolor=black]{hyperref}

\newcommand{\black}[1]{{\color{black} {#1}}}

\usepackage{verbatim}
\usepackage{color}

\begin{document}

\title{\textit{Ab initio} investigation of {laser-induced} ultrafast demagnetization of L1$_0$ FePt: {Intensity dependence and importance of electron coherence}}

\author{M. S. Mrudul}
\affiliation{%
Department of Physics and Astronomy, P.\,O.\ Box 516, Uppsala University, S-75120 Uppsala, Sweden}

\author{Peter M. Oppeneer}
\affiliation{%
Department of Physics and Astronomy, P.\,O.\ Box 516, Uppsala University, S-75120 Uppsala, Sweden}

\date{May 26, 2023}
\begin{abstract}
We  theoretically investigate the  optically-induced demagnetization of ferromagnetic FePt using the time-dependent density functional theory (TDDFT). We compare the demagnetization mechanism in the perturbative and nonperturbative limits of light-matter interaction and show how the underlying mechanism of the ultrafast demagnetization depends on the driving laser intensity. Our calculations show that the femtosecond demagnetization in TDDFT is a longitudinal magnetization reduction and results from a nonlinear optomagnetic effect, akin to the inverse Faraday effect.  The demagnetization scales quadratically with the electric field $E$ in the perturbative limit, i.e., $\Delta M_z \propto E^{2}$. Moreover, the magnetization dynamics happens dominantly at even multiples $n\omega_0$, ($n = 0, 2, \cdots$) of the pump-laser frequency $\omega_0$, whereas odd multiples of $\omega_0$  do not contribute. We further investigate the demagnetization in conjunction to the optically-induced change of electron occupations and electron correlations. 
Correlations within the Kohn-Sham local-density framework are shown to have an appreciable yet distinct effect on the amount of demagnetization depending on the laser intensity. 
Comparing the \textit{ab initio} computed demagnetizations with those calculated from  spin occupations, we show that electronic coherence plays a dominant role in the demagnetization process, whereas interpretations based on the time-dependent occupation numbers poorly describe the ultrafast demagnetization.
\end{abstract}
\maketitle

\section{Introduction} 
A number of experiments in the late 1990s, pioneered by Beaurepaire \textit{et al.}, unanimously unveiled the possibility of ultrafast laser-induced demagnetization of metallic ferromagnets~\cite{beaurepaire1996ultrafast,hohlfeld1997nonequilibrium,scholl1997ultrafast}. 
Ultrafast demagnetization received tremendous attention due to {the prospects of manipulating}  magnetic moments orders of magnitude faster than {then-known} conventional techniques. This remarkable discovery brought forth {the field of} femtomagnetism \cite{Carva2017,Scheid2022}, {the development of} which could benefit ultrafast spintronics and faster magnetic data storage devices~\cite{tudosa2004ultimate}. The field of femtomagnetism is currently well established with recent additions of laser-induced magnetic phase transition~\cite{ju2004ultrafast}, coherent control of antiferromagnetic spin waves~\cite{kampfrath2011coherent}, and  all-optical magnetization switching~\cite{stanciu2007all}, to name a few.

Apart from its vast technological applications, the physics underlying fast demagnetization continues to {fascinate}
researchers. It is one process that gives access to a time-domain understanding of scattering mechanisms between electrons, spins and phonons. However, there are several fundamental issues yet unresolved in ultrafast demagnetization, such as the transfer of angular momentum, interpretation of the magneto-optical signals of nonthermal electrons, and the \black{entangled  mechanisms} of ultrafast demagnetization \black{on short and longer time scales}~\cite{kirilyuk2010ultrafast,Carva2017,Lloyd-Hughes2021}.

The demagnetization mechanism is initialized by an electronic excitation induced by the external laser field, {with typically an 800-nm wavelength}. This is followed by the thermalization of electrons, diffusion of spins, and thermal equilibration {through} interactions between electrons, spins, and phonons~\cite{beaurepaire1996ultrafast,battiato2010superdiffusive,Carva2017}. It is nearly impossible to have a single theory incorporating all these processes happening at different time and length scales. However, the initial step can be assumed to be governed by {the} electron-photon interaction, at least for a short laser pulse of a few femtoseconds duration. Moreover, there has to be a non-linear optical process that demagnetizes the material, as there is no linear coupling between spins and photons, except as a relativistic interaction \cite{Mondal2015}.

To understand how the nonequilibrium electron excitation relates to the mechanism of ultrafast demagnetization requires a fully quantum mechanical theory.
Real-time time-dependent density functional theory (TDDFT) \cite{Marques2012} has {recently} been successful in disclosing some of the key aspects of optically-induced demagnetization, including the role of spin-orbit coupling \cite{krieger2015laser}, the role of electron correlations \cite{acharya2020ultrafast,chen2019role,barros2022impact}, the role of noncollinear spins~\cite{chen2019role}, and optical intersite spin-transfer (OISTR) \cite{dewhurst2018laser,Willems2020,hofherr2020ultrafast,Tengdin2020,elhanoty2022element}.

In the present work, we addresses how {essentially} nonlinear optical excitations can induce ultrafast demagnetization, employing TDDFT simulations. We use the L1$_0$-ordered phase of the iron platinum alloy (L1$_0$ FePt) for this study. This material is a strong candidate for magnetic recording due to its high magneto-crystalline anisotropy~\cite{weller2013l10,challener2009heat} {and it has been thoroughly investigated in recent theoretical and experimental studies
\cite{Yamamoto2019,maldonado2017theory,Reid2018,Yamamoto2019,Sharma2021}.}
Moreover, FePt is a two-component ferromagnet which allows us to explore also the role of inter-site excitations.

\section{Methodology}

We performed real-time TDDFT  calculations to understand the microscopic mechanism of ultrafast demagnetization. We used exchange-correlation functionals within local spin density approximation (LSDA)~\cite{perdew1981self}. Our TDDFT calculations within adiabatic LSDA are performed on a real-space grid, as implemented in the Octopus package \cite{tancogne2020octopus}. Fully relativistic, norm-conserving Hatwigsen-Goedecker-Hutter (HGH) pseudo-potentials were employed, for which relativistic corrections are implemented as a nonlocal operator.

We solved time-dependent Kohn-Sham equation (TDKS) to obtain the laser-induced changes in the Kohn-Sham Bloch state of the $n$-${\rm th}$ band at $\bm{k}$ in the Brillouin zone. The TDKS equation in atomic units reads,
 \begin{equation}
 \begin{split}
 \! i\frac{\partial}{\partial t}\left|\psi_{n\bm{k}}(\bm{r},t)\right\rangle &= \Biggl\{\frac{1}{2}\left(-i \bm{\nabla}+\frac{1}{c}\bm{A}(t)\right)^2  +v_{\rm ext} + v_{\rm nl}\Biggr.\\
 &\Biggl.+v_{\rm H}[n(\bm{r},t)]+v_{\rm xc}[n(\bm{r},t)] \Biggr\}\left|\psi_{n\bm{k}}(\bm{r},t)\right\rangle.
 \end{split}\label{eq:TDKS}
 \end{equation}
Here, $\left|\psi_{n\bm{k}}(\bm{r},t)\right\rangle$ is a Pauli spinor  with the inclusion of relativistic corrections. The laser field is incorporated through the vector potential $\bm{A}(t)$, $v_{\rm ext}$ is the external potential representing the electron-ion interaction, and $v_{\rm nl}$ is the nonlocal part of the potential accounting for relativistic corrections. $v_{\rm H}$ and $v_{\rm xc}$ are the Hartree and exchange-correlation potentials, respectively. In the present work, wave functions are time propagated with a time step of 4.8 attoseconds.

To comprehend the importance of time-dependent electron correlations,  we have also time-evolved the Kohn-Sham wave functions freezing $v_{{\rm H}}$ and $v_{\rm xc}$ to their ground state values. This is called an independent particle (IP) approach, as the electrons are excited in the ground-state band-structure, irrespective of how other electrons are time-evolved~\cite{mrudul2020high,chen2019role}. Note that with correlations we mean here those within the Kohn-Sham DFT framework, not an additionally added electron Coulomb interaction. 

The lattice parameters of the primitive tetragonal unit cell of L1$_0$ FePt are $a$, $b$ = 2.72 \AA$ $, and $c$ = 3.76 \AA$ $~\cite{maldonado2017theory}. We sampled the Brillouin zone with a 9 $\times$ 9 $\times$ 9 grid and the unit cell with a uniform spacing of 0.13 {\AA}. The magnetic ground state is in reasonable agreement with previous works, with magnetic moments of 2.87~$\mu_B$ (Fe) and 0.29~$\mu_B$ (Pt) along the $c$ axis~\cite{maldonado2017theory,oppeneer1998magneto}.

The vector potential of the laser field is modeled with a sin$^2$ envelope. In this case, the fluence of a laser pulse with peak intensity $I_0$ and duration $\mathcal{T}$ is obtained as 3$I_0\mathcal{T}$/8. Throughout the paper, we use a laser pulse linearly polarized along the $x$ axis with a carrier wavelength of 800 nm ($\hbar \omega_0 = 1.55 $ eV) and a pulse duration of 15 fs. The magnetization mentioned in the manuscript is spin magnetization unless stated otherwise. Also, {as} the change in magnetization is observed only along the $z$ axis, 
other components are {not further considered here.} 

\section{Results}
\begin{figure}
	\centering
	\includegraphics[width=\linewidth]{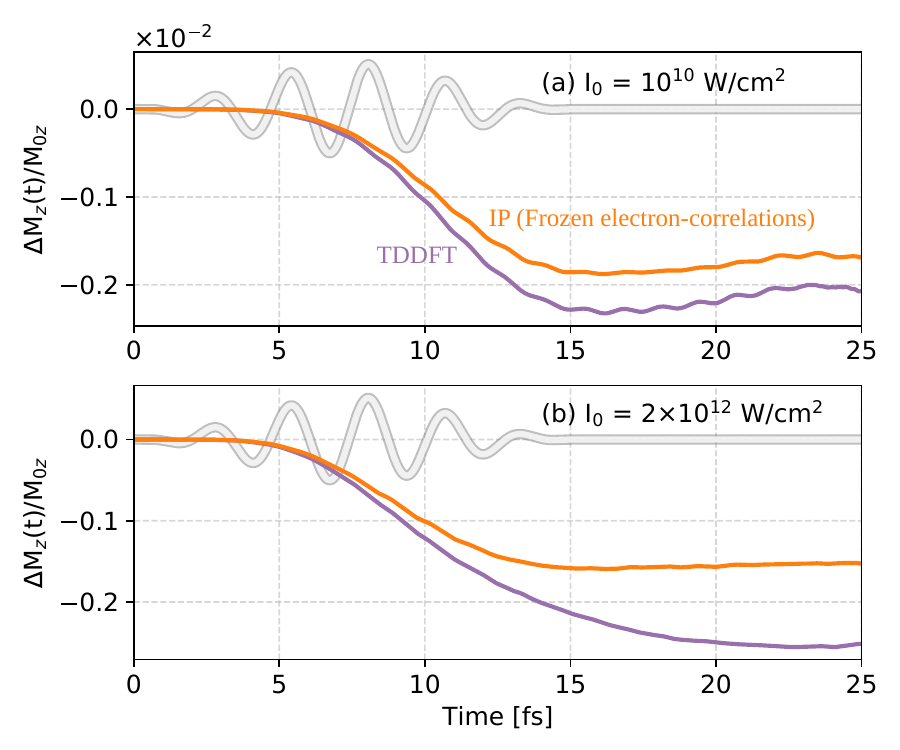}
	\caption{Laser-induced change in the magnetization (scaled) of L1$_0$ FePt for incident laser intensities (fluences) of (a) 10$^{10}$ W/cm$^2$ (56.25 $\mu$J/cm$^2$), and (b) 2$\times$10$^{12}$ W/cm$^2$ (11.25 mJ/cm$^2$), calculated within TDDFT (violet) and IP (orange) approaches. The vector potential of the laser field (grey) has a central wavelength of 800 nm.}
	\label{fig:fig1}
\end{figure}

In Fig.~\ref{fig:fig1}, we present the change in magnetization calculated for L1$_0$ FePt with laser pulses of peak intensities 10$^{10}$ W/cm$^2$ (Fig.~\ref{fig:fig1}(a)) and 2$\times$10$^{12}$ W/cm$^2$ (Fig.~\ref{fig:fig1}(b)). Their respective fluences are 56.25 $\mu$J/cm$^2$ and 11.25 mJ/cm$^2$. These two laser pulses are respectively referred to as `weak' and `strong' throughout the paper.

We observe that the change in magnetization starts after a few femtoseconds of the laser interaction. Moreover, the order of magnitude of demagnetization scales with peak intensity. A noticeable demagnetization close to 25\% is achieved for the strong pulse (Fig.~\ref{fig:fig1}(b)), {while for the weak pulse it is about 0.23\%}. Interestingly, {the} magnetization changes even after the laser pulse. This change for the weak pulse is like a long wavelength oscillation, whereas the material continues to demagnetize for a few {more} femtoseconds for the strong pulse. 

At this point it is instructive to remark that, propagation effects are not in the present calculations. Thus, the laser fluence is completely absorbed in the unit cell and is not the same as the pump fluence outside of the material, as usually denoted in experiments. Specifically,
{the energy transferred per unit cell of the material for the weak and strong laser pulses are 23 meV and 3 eV, respectively.} {The latter value is significantly higher than typical experimental values.}

Comparing results from the TDDFT and IP approaches reveals that the electron correlation enhances demagnetization
{(see Fig.\ \ref{fig:fig1})}. This indicates the strong influence of time-dependent electron correlations in the microscopic mechanism of ultrafast demagnetization. A similar conclusion {was} obtained for Ni in Ref.~\cite{chen2019role}. In addition, the importance of choosing a better exchange-correlation functional beyond LSDA~\cite{barros2022impact}, and the significance of going beyond {the} adiabatic approximation in TDDFT~\cite{acharya2020ultrafast} were discussed recently.

In the following subsections, {we perform} a systematic investigation, {to} show how the possible mechanism of ultrafast demagnetization in TDDFT is related to an optomagnetic effect, the inverse Faraday effect \cite{Vanderziel1965}. Furthermore, we demonstrate how the role of electron correlations is different in weak and strong intensity regimes. Finally, we discuss additional contributions to the element-resolved magnetization dynamics and highlight the role of coherence in the demagnetization process, in contrast to the optically-induced changes of  electron occupations. 

\subsection{Nonlinear mechanism of the demagnetization}
\label{section:raman}
\begin{figure}
	\centering
	\includegraphics[width=\linewidth]{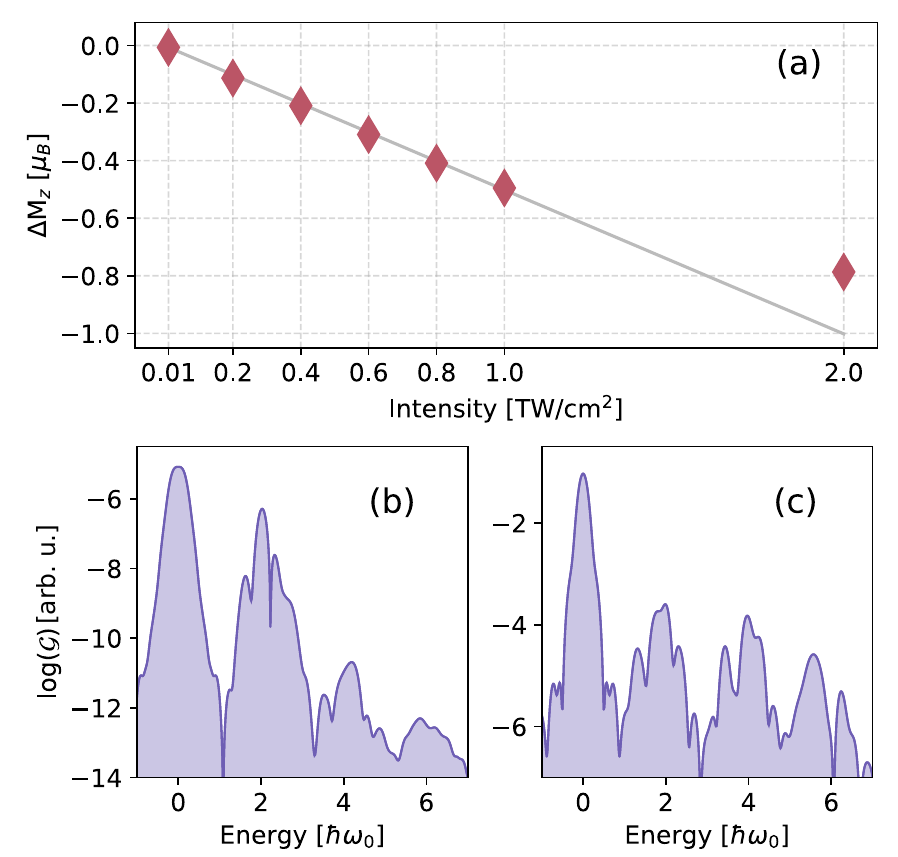}
	\caption{(a) Laser-induced change in magnetization as a function of the peak intensity of the laser field. Magnetization dynamics in the frequency domain ($\mathcal{G}$, see text for definition), for incident laser intensities of (b) 10$^{10}$ W/cm$^2$ and (c) 2$\times$10$^{12}$ W/cm$^2$.}
	\label{fig:scale1}
\end{figure}	

We presented in Fig.~\ref{fig:fig1} how the order of magnitude of demagnetization scales with the laser intensity. In Fig.~\ref{fig:scale1}(a), we show how demagnetization depends on the peak intensity of the laser field. The change in the magnetization is estimated at the end of the laser pulse. In Fig.~\ref{fig:scale1}(a), we see a linear dependence of $\Delta M_z$ on laser intensity up to an 
intensity order of 10$^{11}$ W/cm$^2$. However, there is an apparent deviation from this behavior in a stronger laser field.

To understand {the} magnetization dynamics in the frequency domain, we define $\mathcal{G} = |\mathcal{FT}(\partial M_z/\partial t)|^2$, where $\mathcal{FT}$ is the Fourier transform. Figures ~\ref{fig:scale1}(b) and \ref{fig:scale1}(c) present calculated $\mathcal{G}$ for the weak and the strong laser fields, respectively. The figures 
{reveal} that the change in magnetization happens at even orders of the {pump} laser frequency. For a material with inversion symmetry, the dynamical symmetry of the time-dependent Hamiltonian allows {the} magnetization to follow this behavior (see Appendix~\ref{appendix:A} for the proof).

\begin{figure}
	\centering
	\includegraphics[width=\linewidth]{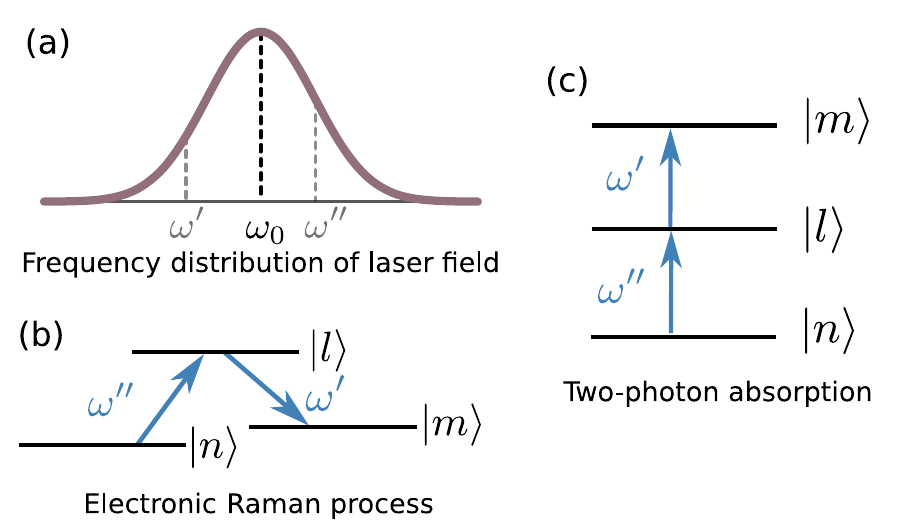}
	\caption{(a) A typical frequency distribution of an ultra-short laser field with central frequency $\omega_0$. (b) Electronic Raman excitation, and (c) two-photon absorption, where $|n\rangle$,  $|l\rangle$, and $|m\rangle$ are electronic states at a particular $\bm{k}$ point.}
	\label{fig:diagram}
\end{figure}

For the weak laser, the prominent magnetization dynamics happens close to zero and 2$\omega_0$ frequencies (Fig.~\ref{fig:scale1}(b)). Let us try to understand the possible mechanisms of magnetization dynamics in the weak/perturbative limit of laser interaction. The typical frequency distribution of an ultrashort laser pulse with central frequency $\omega_0$ is shown in Fig.~\ref{fig:diagram}(a). The figure shows that a range of frequencies centered around $\omega_0$ co-exist in the laser field. From Figs.~\ref{fig:scale1}(a) and \ref{fig:scale1}(b), we can deduce that the change in magnetization is due to a mechanism described as 
\begin{equation}
\Delta M_i (\omega^\prime\pm\omega^{\prime\prime})  = \alpha_{ijk}E_j(\omega^\prime)E^{*}_k(\omega^{\prime\prime}) + c.c.,
\label{eq:mechanism}
\end{equation}
where $\alpha_{ijk}$ is the optomagnetic susceptibility, $\bm{E}(t)$ is the electric field of the laser, and $\omega^\prime$, and $\omega^{\prime\prime}$ are frequencies of the laser field.

It is well known that there is no linear coupling term between spins and photons within the dipole approximation. This {implies} that the mechanism of light-induced change in magnetization is a nonlinear optical process involving at least two photons. Schematic diagrams of two possible electronic processes  that satisfy Eq.~(\ref{eq:mechanism}) are shown in Figs.~\ref{fig:diagram}(b) and ~\ref{fig:diagram}(c). {Let us} consider different electronic states at a particular $\bm{k}$ point in the Brillouin zone. Intuitively, one of the processes is when an electron in state $|n\rangle$ jumps to another state $|m\rangle$, close in energy, with an intermediate higher energy state $|l\rangle$ (see Fig.~\ref{fig:diagram}(b)). This is the electronic Raman scattering when  $|m\rangle \neq |n\rangle$, otherwise called electronic Rayleigh scattering. These mechanisms contribute to magnetization dynamics close to zero frequency. It is 
{interesting} to note that the electronic Raman process was proposed to create a laser-induced effective magnetic field resulting in the inverse Faraday effect~\cite{kirilyuk2010ultrafast,Vahaplar2012}. However, this theory was {obtained} from thermodynamic energy considerations for circularly-polarized laser {light} interacting on nonabsorbing media. Recently, an \textit{ab initio} material-specific theory confirmed the role of these electronic processes in the inverse Faraday effect~\cite{Popova2011,battiato2014quantum,berritta2016ab}.

The difference between the causes of the inverse Faraday effect and ultrafast demagnetization is an additional contribution from the two-photon absorption (Fig.~\ref{fig:diagram}(c)). The two-photon absorption is responsible for the magnetization dynamics close to 2$\omega_0$ in Fig.~\ref{fig:scale1}(b). Note that, for a centrosymmetric material, electric-dipole transitions are observed between electronic states of opposite parity. In contrast, two-photon interactions such as the electronic Raman or two-photon absorption happen between states of the same parity. So, the selection rules between initial and final states for these processes are similar to electric quadrupole or magnetic dipole transitions~\cite{elliott1963possible}. Moreover, these electronic processes depend on having a range of frequencies in the laser field, pointing out {their} relation to the {ultrashort} duration of the laser pulse.
 
One of the advantages of real-time TDDFT is that it does not restrict us to the perturbative limit of light-matter interaction. {Hence}, higher-order processes such as four-photon absorption are included by default in the calculation. Moreover, all these higher-order interactions have their Raman and Rayleigh counterparts, all contributing near the zero frequency. However, when the laser pulse is weak, we can see that the higher-order interaction terms are {much} weaker, typical {for} the perturbative limit ({note the logarithmic scale in} Fig.~\ref{fig:scale1} (b)). In contrast, in Fig.~\ref{fig:scale1} (c), we can see that second and fourth-order dynamics are happening with comparable intensity. This shows the highly nonperturbative nature of the underlying {high-intensity} mechanism. These nonperturbative contributions are responsible for the deviation from the linear intensity dependence in Fig~\ref{fig:scale1}(a). In short, we {have} demonstrated the {essential} sub-femtosecond processes of ultrafast laser-induced demagnetization.

Previous calculations of the light-induced change of the magnetization through the inverse Faraday effect -- which can happen even with linearly polarized light in a ferromagnet -- gave values $10^{-2} - 10^{-3}$ $\mu_B$ per atomic volume for (continuous wave) laser intensities of $10^{10}$ W/cm$^2$ \cite{berritta2016ab}, quite comparable with the TDDFT result.

\subsection{Insights from out-of-equilibrium spin-dependent occupations}
\label{Sec:3B}

\begin{figure*}[t!]
	\centering
	\includegraphics[width=\linewidth]{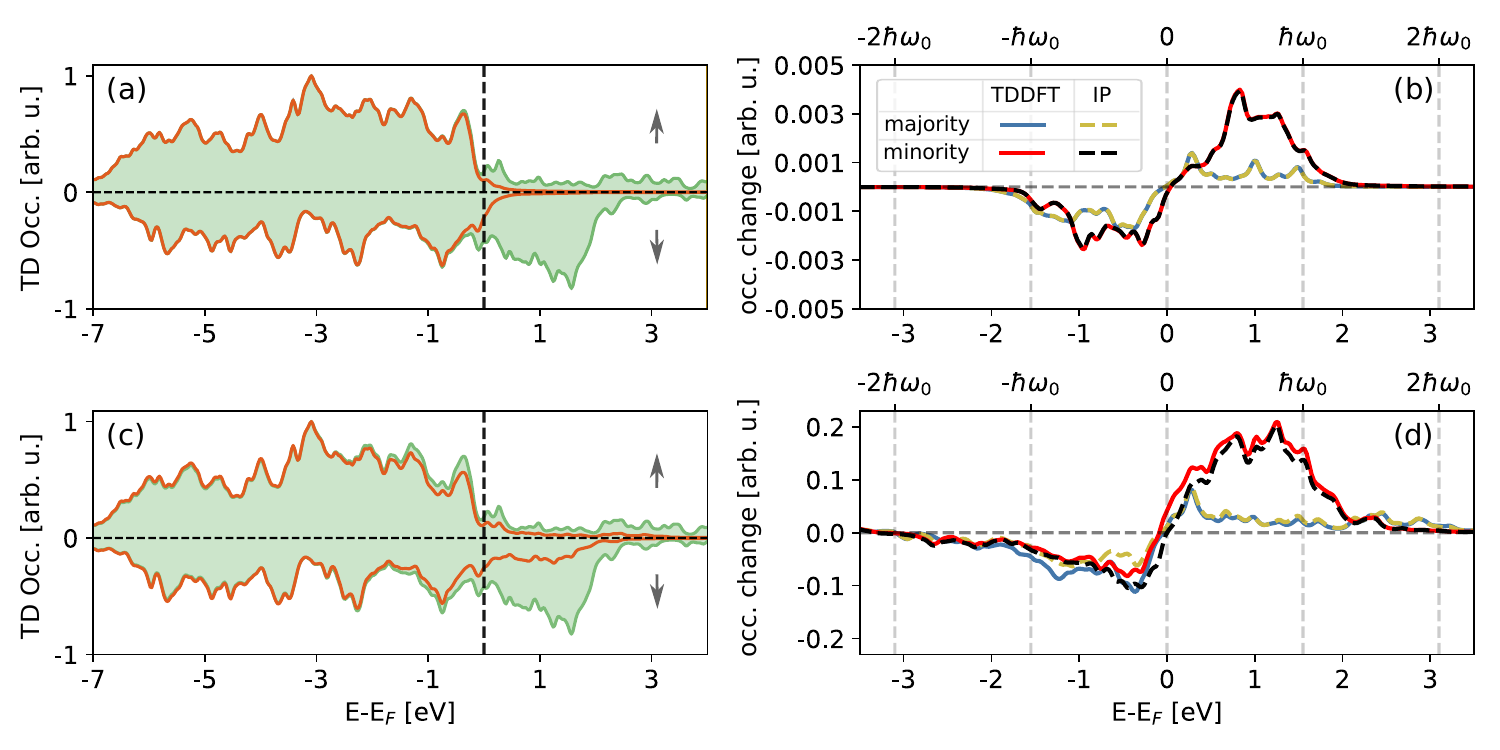}
	\caption{Laser-induced {modification of} electron occupations (orange line) {after the laser pulse} for a laser of peak intensity (a) 10$^{10}$ W/cm$^2$, and (c) 2$\times$10$^{12}$ W/cm$^2$. Majority and minority electrons are represented, respectively, {by the} positive and negative $y$ axis. The ground-state density of states is shown {by the} green {shaded area}. {Panels (b) and (d) show the} laser-induced change in the electron occupations corresponding to panels (a) and (c), respectively. }
	\label{fig:tddos}
\end{figure*}

\begin{figure}
	\centering
	\includegraphics[width=\linewidth]{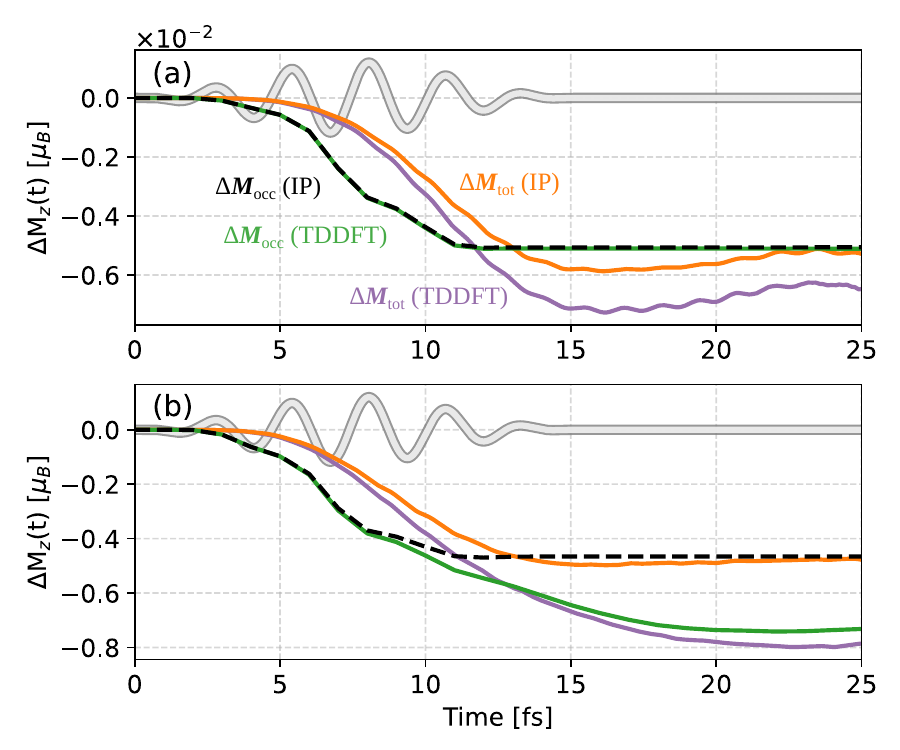}
	\caption{Ultrafast laser-induced demagnetization {computed with TDDFT} ({$\Delta \bm{M}_{\rm tot}$}) along with the magnetization change due to change in the electronic occupations ($\Delta \bm{M}_{\rm occ}$) for a laser of peak intensity (a) 10$^{10}$ W/cm$^2$, and (b) 2$\times$10$^{12}$ W/cm$^2$. {For comparison, results computed with the IP approximation are also given.}}
	\label{fig:spinocc}
\end{figure}

In this section, we will explore the role of band structure, {spin-dependent occupation numbers}, electron correlation, {and coherence} in the mechanism of ultrafast demagnetization through the 
{formalism} of density matrices.

It is 
important to mention that spin-orbit coupling is at the heart of ultrafast demagnetization in TDDFT \cite{krieger2015laser} also of  these nonlinear optomagnetic processes \cite{berritta2016ab}. Without spin-orbit coupling, the spin can obviously not be changed. 
We start therefore with considering the Kohn-Sham Bloch state, $\psi_{n\bm{k}}(\bm{r})$, {which} is a linear superposition of pure spin states given by, $\psi_{n\bm{k}}(\bm{r}) = [a_{n\bm{k}}(\bm{r})| \!\!\uparrow\rangle + b_{n\bm{k}}(\bm{r})| \!\! \downarrow\rangle] e^{i\bm{k}\cdot\bm{r}}$. For majority-spin, the term $b_{n\bm{k}}$ describes the spin mixing in the Bloch state. Consider a dipole allowed transition between two of such states separated by an energy of $\hbar\omega_0$. The dipole transition of an electron between these states could change the spin polarization due to the different {amounts of} spin mixing present in these states. This is analogous to the Elliott model for spin loss due to electronic transitions \cite{Elliott1954}, which has been considered previously \cite{Steiauf2009,Krauss2009,Carva2013,Baral2016}.  This mechanism is referred to as demagnetization due to light absorption in Ref.~\cite{scheid2019ab}. However, this needs to be reconciled with
the idea that {electric} dipole transition should preserve magnetization. {Below}, we will show that a change in spin \black{mixing} {and the related spin occupation numbers}
does not necessarily imply a change in magnetization during the laser interaction.

{Next, we introduce the density matrix.} The {set of} excited Kohn-Sham wave functions, $\left\lbrace |\psi_{n\bm{k}}(t)\rangle\right\rbrace$ can be expanded in terms of the ground-state Kohn-Sham wave functions, $\left\lbrace |\psi^0_{n\bm{k}}\rangle\right\rbrace$, as $|\psi_{n\bm{k}}(t)\rangle = \sum_{m}\alpha_{m}^{n\bm{k}}(t)|\psi^0_{m\bm{k}}\rangle$. The time-dependent density matrix elements between states $|\psi^0_{i\bm{k}}\rangle$ and $|\psi^0_{j\bm{k}}\rangle$ can be defined as 
\begin{equation}
\rho_{ij}^{\bm{k}}(t) = \sum_n f_{n\bm{k}}\alpha_i^{n\bm{k}}(t)\alpha_j^{n\bm{k}*}(t),
\end{equation}
where $f_{n\bm{k}}$ is the initial occupation of $|\psi^0_{n\bm{k}}\rangle$. Here, the diagonal elements of the density matrix elements govern the time-dependent occupations, and off-diagonal elements describe the electron coherence between the states.

The nonequilibrium electron occupations can be estimated as, {$g (\epsilon )= \sum_{n,\bm{k}} \rho_{nn}^{\bm{k}}\delta(\epsilon_{n\bm{k}}-\epsilon) $}, where $\epsilon_{n\bm{k}}$ is the energy eigenvalue of the state $|\psi^0_{n\bm{k}}\rangle$. {In} Figs.~\ref{fig:tddos}(a) and \ref{fig:tddos}(c) {we show} 
the electron occupations after the laser pulse for the weak and the strong laser fields.
The green shaded area in Fig.~\ref{fig:tddos} {depicts} the density of states of the electrons.

The nonlinear mechanisms described in Sec.\ \ref{section:raman} guarantee only a change in the magnetization. It is the peculiar shape of the density of states in ferromagnets that guides demagnetization. It is clear from the density of states that any nonequilibrium spin occupation tends to demagnetize the material. This is also true even if the laser field has a helicity~\cite{berritta2016ab} unlike the inverse Faraday effect in nonmagnetic materials.

For getting a better picture, the spin-resolved change in occupations after the laser pulse is plotted in Figs.~\ref{fig:tddos}(b) and \ref{fig:tddos}(d). For the weak pulse, optical transitions happen for electronic states between $\pm \hbar \omega_0$. On the other hand, when the laser pulse is strong, transitions happen {even} in the range $\pm 2\hbar \omega_0$. 
In both cases, an increase in the number of minority electrons and depletion in the number of majority electrons can be recognized. 

The contrasting difference in the weak and strong laser regimes observed here is in the role of electron correlations. When the laser pulse is weak, the excited electron occupations are the {identical} for TDDFT and the IP approximation. This is surprising since the same is not true for the demagnetization shown in Fig.~\ref{fig:fig1}(a), where electron correlations enhance demagnetization. On the other hand, for a stronger laser pulse, dynamical electron correlations 
increase the {amount of} spin-flip excitations.

To better understand the time-evolution of the electrons, we analyze the equation of motion of the density-matrix elements. 
The time-dependent Kohn-Sham Hamiltonian can be effectively written as, $\hat{h}(t) = \hat{h}_0 + \delta h_{ee}(t) + \hat{h}_{lm}(t)$. Here, $\hat{h}_0$ is the ground-state Hamiltonian. The time dependence in the Hamiltonian is attributed to dynamical changes in the electron-electron potentials, $\delta h_{ee}(t)$, and the laser-matter interaction, $\hat{h}_{lm}(t)$. The difference between the IP and TDDFT approaches is the presence of {the} $\delta h_{ee}(t)$ term in the Hamiltonian. {The} equation of motion for the density-matrix elements is given by,
\begin{equation}
\label{eq:dmat}
	\frac{\partial}{\partial t} \rho_{nn'} = -i \left\{ \rho_{nn'}(\epsilon_{n}-\epsilon_{n'})  - [\hat{\rho},\delta\hat{h}_{ee} +\hat{h}_{lm}]_{nn'} \right\}. 
\end{equation}
Note that these excitations happen between different states at a particular $\bm{k}$ point, and the $\bm{k}$-index is implicitly included.  The magnetization dynamics follows as,
\begin{equation}
	\label{eq:mag}
	\frac{\partial}{\partial t} \bm{M} \! = -i \sum_{nn'} \! \left\{ \bm{M}_{n'n}\rho_{nn'}(\epsilon_{n}-\epsilon_{n'}) + \bm{M}_{n'n}[\hat{\rho},\delta\hat{h}_{ee}]_{nn'}\right\}\!,
\end{equation}
where $\bm{M}_{nn'}$ are the matrix elements of the magnetization operator.

We could {notice} a few points from Eq.~(\ref{eq:mag}). The role of spin-orbit coupling strength is evident with the presence of $\bm{M}_{nn'}$, which would otherwise be zero between different states. Moreover, as expected, there is no linear dependence of $\hat{h}_{lm}$ and {the} magnetization appearing. 

In Fig.~\ref{fig:spinocc} we compare the change in magnetization $\bm{M}_{\rm tot}$ within the TDDFT and IP approaches with the demagnetization due to the changes in  the occupation numbers.  
We define the magnetization due to a change of spin occupations $\bm{M}_{\bm{occ}}$ as 
\begin{equation}
\bm{M}_{\rm occ}(t) = \frac{1}{N_k}\sum_{n,\bm{k}} \bm{M}^{\bm{k}}_{nn} \rho^{\bm{k}}_{nn} (t),
\end{equation}
where $N_k$ is the number of $\bm{k}$ points. By definition, $\bm{M}_{\rm occ}=\bm{M}_{\rm tot}$ in the ground state.
We can clearly see from Fig.~\ref{fig:spinocc} that $\bm{M}_{\rm occ}$
and $\bm{M}_{\rm tot}$ (same as in Fig.~\ref{fig:fig1}) have different temporal behaviors. 

We start by analyzing the magnetization due to spin-occupation numbers. Figure~\ref{fig:spinocc}(a) shows that $\bm{M}_{\rm occ}$ is the same within TDDFT and IP approaches when the laser pulse is weak, consistent with Fig.~\ref{fig:tddos}(b). However, the behavior of $\bm{M}_{\rm occ}$ has considerable deviation close to the end of the laser pulse for the strong pulse (Fig.~\ref{fig:spinocc}(b)). From Eq.~(\ref{eq:dmat}), it is evident that, after the laser pulse, modifications in electron occupations are governed by the change in electron correlations. Thus, $\bm{M}_{\rm occ}$ is {expected} to be a constant after the laser pulse when calculated within the IP approach, in agreement with Fig.~\ref{fig:spinocc}. The role of electron correlation for the strong laser pulse is in increasing the number of excited electrons, resulting in a higher demagnetization. In addition, electron-correlation-driven modifications in $\bm{M}_{\rm occ}$ continue to demagnetize the material five more femtoseconds after the laser pulse. It is essential to mention that the mechanism of demagnetization for a strong laser pulse is beyond a static band-structure picture.

When an electron is excited by light, an electronic coherence is formed between states in addition to the transfer of occupations. As mentioned earlier, electronic coherence is included in the off-diagonal density-matrix elements. So, the difference between $\bm{M}_{\rm occ}$ and $\bm{M}_{\rm tot}$ is this contribution. It is evident that the first driving term in Eq.~(\ref{eq:mag}) is entirely governed by electronic coherence.
For example, when a two-photon absorption modifies the magnetization, the magnetization dynamics will have a frequency component of 2$\omega_0$. On the other hand, the long wavelength oscillations observed in Fig.~\ref{fig:fig1}(a) are due to the electronic coherence contribution of electronic Raman excitations. Clearly, in Fig.~\ref{fig:spinocc}(a), the role of electron-electron interaction is to enhance the electron coherence between the states keeping the occupations of electrons fixed. We explain below how this is related to the electronic Raman mechanism.

Electronic Raman excitations do not have to change the occupation of states. Suppose $|m\rangle$ and $|n\rangle$ are two partially occupied states. Consider the case when the electronic Raman process happens between these states simultaneously back and forth. This means that there is no transfer of occupation between these states, but coherence is induced. Such a process requires two electrons at different electronic states to communicate. This is why the coherence contribution increases during the laser interaction with the support of dynamic modification of electron correlations.

Additionally, we note that the change in  $\bm{M}_{\rm occ}$ starts even before the total magnetization $\bm{M}_{\rm tot}$ begins to demagnetize. The demagnetization due to a change of the spin occupations has already reached about 75\% at the middle of the pump pulse, whereas $\bm{M}_{\rm tot}$ just begins to reduce. The rapid change of the spin occupations can be attributed 
to dipole excitations between spinor states with different spin mixing, as discussed in the beginning of this section.
However,
such excitations conserve clearly the total magnetization, compensated by electron coherence.     
 
Note that the relative contribution from the coherence is weaker for a strong laser pules. In addition, persistence of these effects is characterized by the dephasing time, which is usually shorter than the thermalization time of electrons.  

Our findings are furthermore supported by the spectral distribution of excited Kohn-Sham wave functions presented in Appendix~\ref{appendix:B}.
These show that electronic transitions happen at $\pm \hbar\omega_0$ for the weak laser intensity, while  some transitions occur also at $\pm 2\hbar \omega_0$ for the strong laser intensity. 
The magnetization-changing electronic Raman transitions near $\hbar \omega \approx 0$ increase notably with laser intensity as well as with the dynamical modification in the electron correlations, giving a stronger demagnetization.

In short, our results show how the role of electron correlation in ultrafast demagenetization is distinct depending on the laser intensity. In the strong laser field, electron correlation modifies the electron occupation substantially, even after the laser pulse, resulting in a stronger demagnetization. On the other hand, in the weak field, electron correlation enhances the electronic Raman excitations and results in an enhanced demagnetization even without changing the occupations. 
This illustrates how the role of electron coherence is crucial in understanding these processes and proves that a change in spin occupation does not necessarily imply a change in magnetization.

\subsection{Element-resolved contributions}

\begin{figure*}
	\centering
	\includegraphics[width=\linewidth]{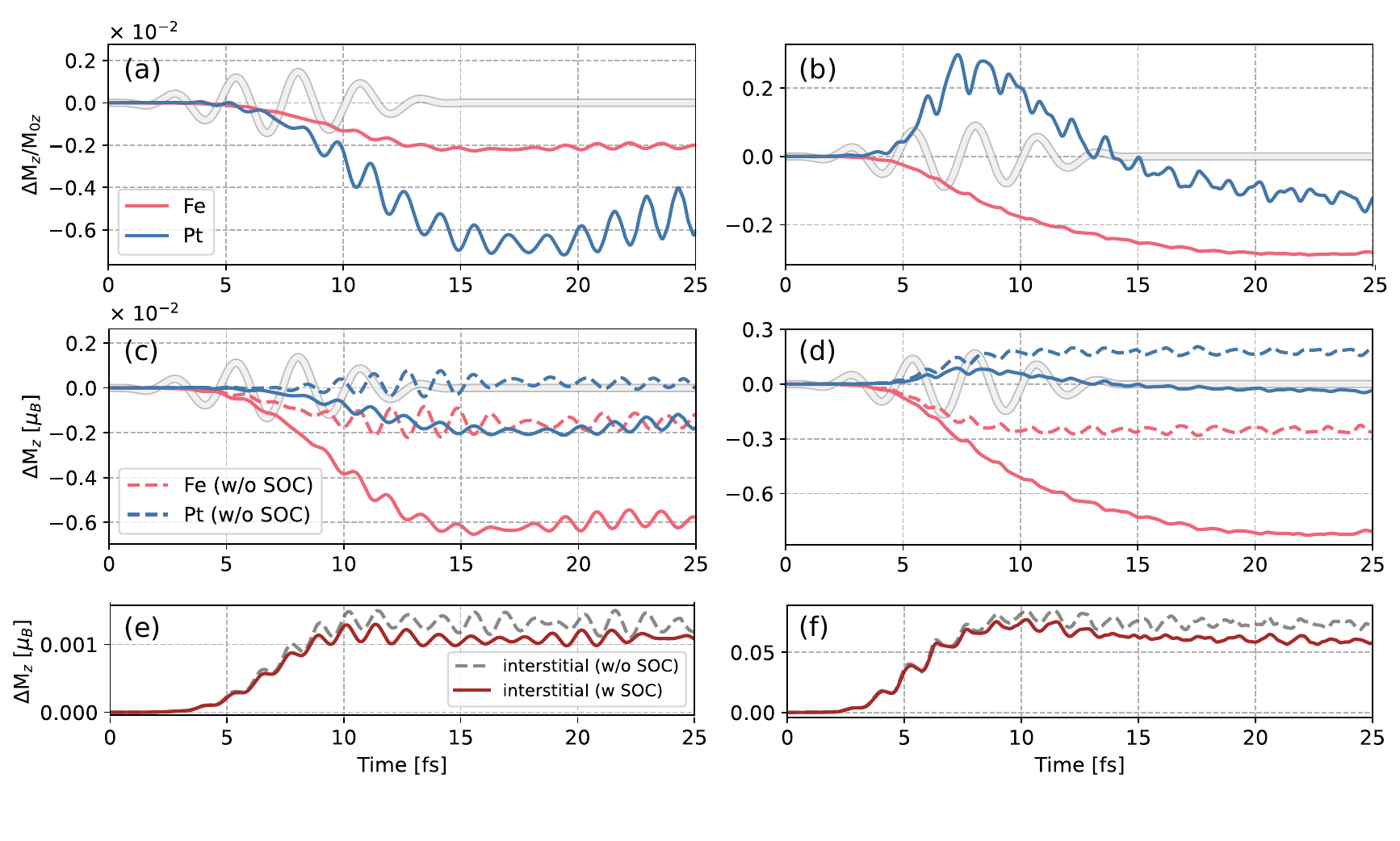}
	\caption{Element-resolved change in magnetization of L1$_0$ FePt for a laser with  a peak intensity of 10$^{10}$ W/cm$^2$ (left panels), and 2$\times$10$^{12}$ W/cm$^2$ (right panels). The demagnetization of Fe and Pt is normalized in panels (a) and (b). The absolute value of the demagnetization is given in panels (c) and (d). Change in the magnetization in the interstitial region is presented in panels (e) and (f).}
	\label{fig:intersite}
\end{figure*}

\begin{figure}
	\centering
	\includegraphics[width=\linewidth]{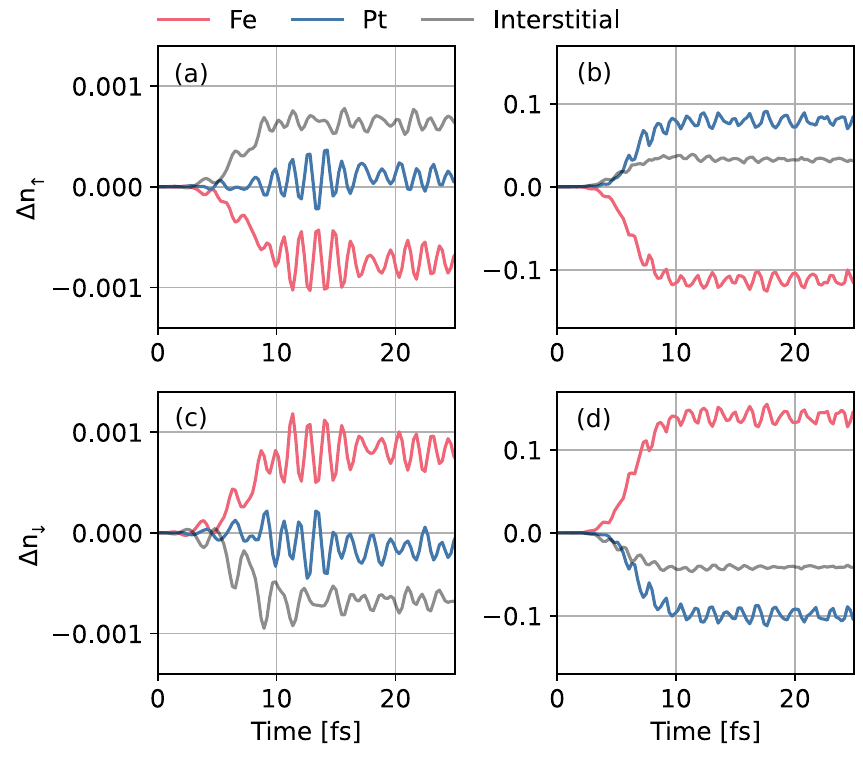}
    	\caption{The change in the number of majority (top panels) and minority electrons (bottom panels) around atoms when lasers of peak intensity 10$^{10}$ W/cm$^2$ (left panels)  and 2$\times$10$^{12}$ W/cm$^2$ (right panels) interact \black{with FePt}. The spin-orbit coupling is turned off in these calculations.}
	\label{fig:nel}
\end{figure}

So far, we have discussed how the net magnetization of the material changes due to the laser interaction. In this section, we explore the element-resolved changes in magnetization when a weak or strong laser interacts with L1$_0$ FePt. For that, the magnetization is calculated  within atomic spheres, and, in addition, the interstitial contribution is analyzed. 

Figure~\ref{fig:intersite} presents the element-resolved change in magnetization for weak (left panels) and strong (right panels) laser pulses. Interestingly, the demagnetization in Pt is larger compared to Fe in the weak laser field [Fig.~\ref{fig:intersite}(a)]. In contrast, Fe demagnetizes comparatively more in the strong external field, and an initial magnetization {increase} is observed on Pt during the pump laser field [Fig.~\ref{fig:intersite}(b)]. 
{The increase of magnetization on Pt and a decrease on Fe has previously been referred to as optically induced spin transfer {(see also \cite{dewhurst2018laser,Willems2020,Tengdin2020})}.}
A fluence dependence of OISTR has been computed for FePd$_3$ alloy in Ref.~\cite{elhanoty2022element}.

To understand the underlying process, we compare the element-resolved demagnetization with and without spin-orbit coupling in Figs.~\ref{fig:intersite}(c) and \ref{fig:intersite}(d). Moreover, {the time-resolved} magnetization in the interstitial region is shown in Figs.~\ref{fig:intersite}(e) and \ref{fig:intersite}(f).

We observe a change in magnetization on Fe and Pt even without the spin-orbit coupling. This can be understood in a similar way as the change of spin polarization due to optical excitation in the Elliott model \cite{Elliott1954}. But instead of spin mixing, we have a Bloch wave function of hybridized Fe and Pt states. Before excitation, the wave function consists of mainly spin-majority electrons localized at the Fe atom. The laser excites electrons to unoccupied states with a different hybridization amount of Fe and Pt states.
Such processes conserve the total spin magnetic moment but delocalize it within the unit cell. The redistribution of electrons to delocalized states also changes the number of spin-polarized electrons within the atomic spheres. This charge redistribution is the background of OISTR.
This 
{optical charge} transfer is strongest near the peak of the laser pulse, and it also induces magnetization in the interstitial part of the unit cell as shown in Figs.~\ref{fig:intersite}(e) and \ref{fig:intersite}(f). In contrast to OISTR, the mechanism of demagnetization occurs within the atomic spheres due to the requirement of spin-orbit coupling. This is also supported by the close resemblance of the interstitial magnetization increase with or without spin-orbit coupling.

To understand the delocalization of spin among different regions of the unit cell, we plot the change in the number of spin-polarized electrons in Fig.~\ref{fig:nel}. The spin-orbit coupling is turned off to 
{examine}
the effect of inter-site excitations. In the weak field limit, we see there is no significant change in the number of electrons around Pt (Fig.~\ref{fig:nel}(a) and \ref{fig:nel}(c)). The charge transfer happens between regions Fe and the interstitial region close to Fe (not shown here). This means there is no charge transfer between Fe and Pt in the weak field case. However, in the strong laser field, we see a decrease of majority-spin and increase of minority-spin electrons on Fe and the opposite on Pt.
{Effectively, the re-distribution in the excited state implies that} majority electrons are transferred from Fe to Pt (Fig.~\ref{fig:nel}(b)), whereas the minority electrons are transferred from Pt to Fe (Fig.~\ref{fig:nel}(d)). Both these effects are comparable in magnitude and give rise to a net optical intersite transfer of spin polarization  from Fe to Pt atomic spheres. This process is however not responsible for the laser-induced demagnetization which requires spin-orbit interaction as well as coherence, as shown in Sec.\ \ref{Sec:3B}.

In short, in addition to the underlying spin-flip mechanism, OISTR is also strongly intensity dependent. This results in dissimilar element-resolved demagnetization dynamics depending upon the laser intensity~\cite{elhanoty2022element}. 
The element-resolved demagnetization is governed by the spin-flip excitations of electronic Raman-type for the weak field. Since {these} spin-flip excitations depend on spin-orbit coupling, Pt demagnetizes more than Fe. When the laser pulse is considerably strong, the spin polarization on Pt  initially increases by OISTR and then demagnetizes due to spin-flip Raman excitations, resulting in a weaker concomitant demagnetization. 

\section{Discussion}

Previously, several mechanisms have been proposed to explain ultrafast optically induced demagnetization
\cite{Carva2017}. Among these are influence of spin-orbit coupling \cite{Zhang2000},
transfer of angular momentum to phonons 
\cite{Koopmans2010,Tauchert2022}, ultrafast magnon generation \cite{Carpene2008,Carpene2015,Turgut2016}, transfer of spin angular momentum to orbital angular momentum \cite{Tows2015},  effect of electronic correlations \cite{Tows2015,acharya2020ultrafast}, superdiffusive spin transport \cite{battiato2010superdiffusive,battiato2012theory},  and OISTR \cite{dewhurst2018laser,siegrist2019light,Willems2020,hofherr2020ultrafast,Tengdin2020}. 

Within the here-applied TDDFT framework, it is not possible to  investigate phonon and magnon quasiparticle scattering as demagnetization channels and neither superdiffusive transport, which occurs on a length scale of several nanometers and a time scale of femtoseconds to picoseconds. Our calculations are consistent with previous work \cite{Zhang2000} and TDDFT calculations that emphasized the importance of spin-orbit coupling \cite{krieger2015laser}. The observed need to account for density-functional electron correlations is furthermore consistent with previous TDDFT investigations \cite{chen2019role,acharya2020ultrafast}.

To investigate the spin-to-orbital angular momentum transfer \cite{Tows2015} we present a comparison of laser-induced changes in the spin and orbital contributions of the magnetization in Appendix~\ref{appendix:C}. We find that there is practically no change of the orbital magnetization, in contrast to the strong reduction of the spin magnetic moment  (see Fig.\ \ref{fig:orb}). This might seem to rule out the proposed transfer to orbital angular momentum \cite{Tows2015},  
but we note that it was pointed out recently that real-time \textit{ab initio} methods violate angular momentum conservation even at the electronic level~\cite{simoni2022conservation}.

OISTR \cite{dewhurst2018laser} appears as a redistribution of spin-polarized electrons due to light absorption, present even without spin-orbit coupling. Our calculations show that there is not only a light-induced change of occupations on the Fe and Pt atoms, but also a comparable change in the interstitial region (see Fig.\ \ref{fig:nel}), consistent with  dipole transitions occurring to more delocalized states. The change in magnetization purely from OISTR can be quantified only when the spin-orbit coupling is turned off. In this scenario, it is easy to see that the estimate is related to the electronic occupations projected on the atomic spheres. This estimate, similar to $M_{\rm occ}$, does not depend on the electron coherence. This is further supported by the fact that the change in magnetization due to $M_{\rm occ}$ and OISTR start at an earlier time than the actual demagnetization (see Fig.\ \ref{fig:spinocc}).  Thus, unlike for the demagnetization, time-dependent electron occupations can be used to interpret OISTR.  As the total demagnetization is  zero during OISTR,  the magnetization  decreases on Fe and increases in the interstitial region and on Pt. Such signature of magnetization on the nonmagnetic atom has been observed recently in compounds consisting of ferromagnetic and nonmagnetic elements \cite{hofherr2020ultrafast,Tengdin2020}, using magneto-optical detection in the extreme ultraviolet (XUV) regime. It should be noted however that the measured optical and magneto-optical signals are significantly modified through the presence of nonequilibrium electron occupations \cite{Lloyd-Hughes2021}. These can lead to an artificial increase in the magneto-optical signal as was shown
for the demagnetization of elemental Ni \cite{Carva2009,Hennes2021}, where OISTR is absent. It is therefore not straightforward to attribute an increase in the magneto-optical signal to OISTR.

Lastly, we can elucidate the material-specific nature of ultrafast demagnetization by comparing that of Fe in L1$_0$ FePt and of body-centered cubic (bcc) Fe. The computed demagnetizations are compared in Appendix~\ref{appendix:D}, Fig.\ \ref{fig:Fe}. The results show that Fe in L1$_0$ FePt is demagnetized ten times more than elemental bcc Fe. This can be attributed to the strong spin-orbit coupling of Pt which assists the electronic Raman-type spin-flip excitations.  Moreover, as the density of states of the two materials is different, the qualitative demagnetization behavior and demagnetization time are also considerably different.

\section{Conclusions}

Employing the TDDFT approach, we have investigated how optically induced demagnetization proceeds in ferromagnetic FePt in the perturbative and nonperturbative limits of light-matter interaction.

Our first key finding is that the demagnetization dominantly occurs at zero-frequency excitations and, to a lesser extent, at even multiples of the pump-laser frequency $\hbar \omega_0$, whereas there is vanishing demagnetization occurring at excitations at odd multiples of $\hbar \omega_0$. The dominant demagnetizing light-matter interaction is identified as an electronic Raman process, akin to the inverse Faraday effect. This is vindicated further by our findings that the  demagnetization happening during the laser pulse is a longitudinal, nonlinear effect that scales as $\Delta M_z \propto E^2$ in the perturbative limit.  
This finding is in addition consistent with the fact that there is no nonrelativistic, linear coupling between the spin and photon field \cite{Mondal2015}. 
Moreover, we find that in the nonperturbative regime the magnetization dynamics due to electronic Raman excitations and OISTR are distinct.

Our second key finding is the importance of electron coherence for the demagnetization process. The significance of electron coherence, expressed by nonzero off-diagonal density-matrix elements, becomes evident when one compares demagnetization to the optically-induced change in spin occupations. 
The temporal evolution of spin occupations is drastically different from that of the magnetization. Moreover, a change in spin occupation caused by the optical excitations between Bloch states of different spin mixing doesn't result in a change in magnetization in the initial phase of the light interaction.  
This observation is relevant for OISTR, which is a redistribution of atomic spin occupations due to the laser pulse, as these do not include electron coherence and, hence, cannot  adequately describe the time evolution of demagnetization. 

Our third finding is the importance of correlations within the Kohn-Sham framework. These have an appreciable influence on the demagnetization and cooperate with the electronic Raman transitions to enhance the demagnetization.

Summarizing, our TDDFT calculations provide insight in how the process of ultrafast laser-induced demagnetization takes place.  We believe that our findings will be helpful for the understanding and interpretation of optical demagnetization measurements on an ultrashort times scale, during  and immediately after the laser excitation.

\appendix
\section{Magnetization {behavior} in the frequency domain}
\label{appendix:A}

Consider the time-dependent Kohn-Sham Hamiltonian described in Eq.~(\ref{eq:TDKS}). For simplicity, we assume that the above Hamiltonian is time-periodic, with the period of the laser pulse, i.e., $\hat{h}(\bm{r},t+\tau) = \hat{h}(\bm{r},t)$. Moreover, the periodicity of the laser field implies that $\bm{A}(t+\tau/2) = -\bm{A}(t)$. This approximation neglects any non-periodic modification in the exchange-correlation terms and assumes a slowly varying envelope for the laser field. 

L1$_0$ FePt has inversion symmetry in its ground state. However, this symmetry is broken when the light-matter interaction term $({\propto}~\bm{A}(t)\cdot\hat{\bm{p}}$) is present.  We recognize that the time-periodic Hamiltonian has the following dynamical symmetry in which one simultaneously applies a space inversion ($\bm{r}\rightarrow-\bm{r}$) and a time translation by half the time period ($t \rightarrow t+\tau/2$). 

We can define a unitary operator $\hat{U}$ related to the above dynamical symmetry as $\hat{h}(\bm{r},t) = \hat{U}\hat{h}(-\bm{r},t+\tau/2)\hat{U}^\dagger$. Similarly, the transformation on the magnetization operator is $\hat{\bm{m}}(\bm{r},t) = \hat{U}\hat{\bm{m}}(-\bm{r},t+\tau/2)\hat{U}^\dagger$. Moreover, using the above transformation in the TDKS equation, we can show that the wave functions transform as $\hat{U}^\dagger \psi(\bm{r},t-\tau/2) = \psi(-\bm{r},t)$.

{From the above relations, we obtain the space-time relation of the magnetization as}
$\bm{m}(\bm{r},t) = \bm{m}(-\bm{r},t+\tau/2)$. {Finally, we get the time dependence of the net magnetization by integrating $\bm{m}(\bm{r},t)$ in the unit cell as} $\bm{M}(t) = \bm{M}(t+\tau/2)$.  The Fourier transform of $\bm{M}(t)$ in orders of the laser frequency, $\omega_0$ = $2\pi/\tau$, can be written as
\begin{equation}
\begin{split}
    \bm{M}(n\omega_0) &= \int \bm{M}(t)e^{in\omega_0 t} dt \\
    &= \int \bm{M}(t+\tau/2)e^{in\omega_0 (t+\tau/2)} dt\\
    &= e^{in\pi}\bm{M}(n\omega_0) \, .
\end{split}
\label{eq:magdyn}
\end{equation}
Thus, the magnetization dynamics driven by light {is} 
due  to electronic processes at even orders of the laser frequency, whereas odd orders do not contribute,  consistent with our TDDFT calculations.

\section{The spectral distribution of excited Kohn-Sham wave functions}\label{appendix:B}
\begin{figure}[t!]
	\centering
	\includegraphics[width=\linewidth]{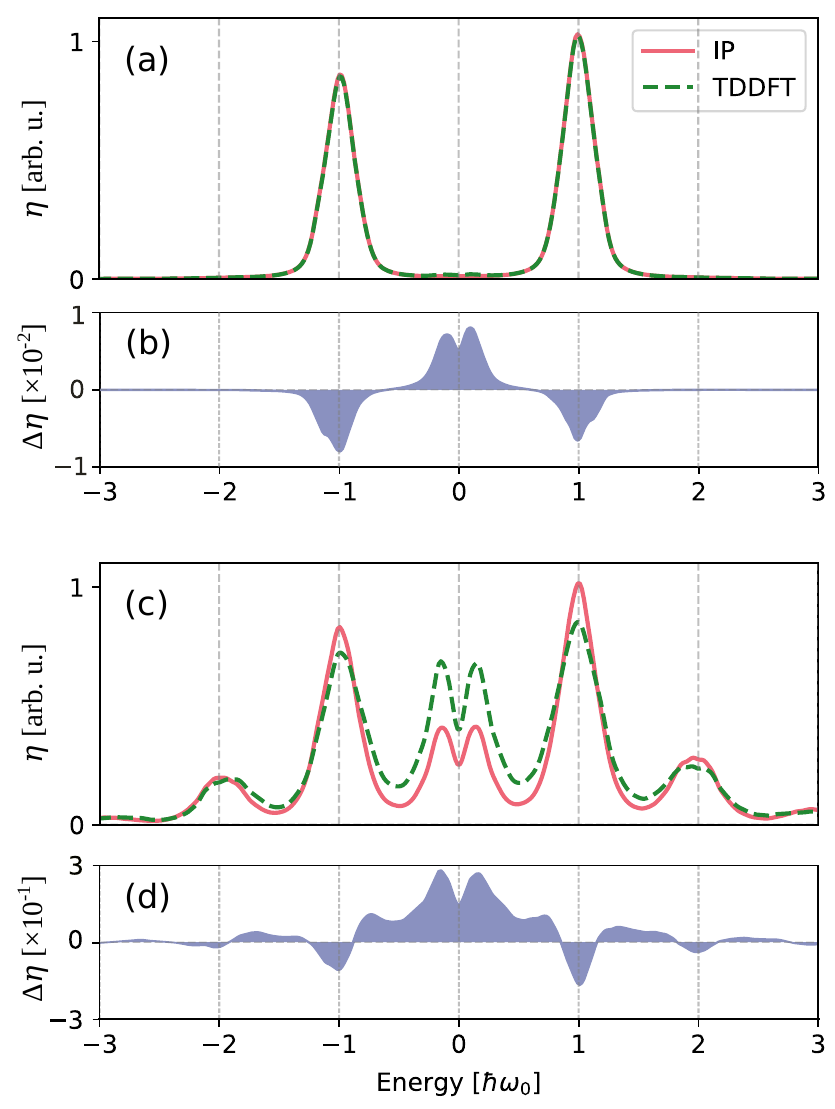}
	\caption{The spectral distribution of excited Kohn-Sham wave function in FePt using the TDDFT and IP approaches for a laser with peak intensity of (a) 10$^{10}$ W/cm$^2$, and (c) 2$\times$10$^{12}$ W/cm$^2$. The difference $\Delta \eta$ ($  = \eta_{\rm TDDFT}- \eta_{\rm IP}$) of the spectral distributions of the two approaches is shown by the blue shaded lines in (b) and (d).}
	\label{fig:spectral}
\end{figure}

The spectral density of the excited state wave function at the end of the laser pulse is defined as, 
\begin{equation}
\eta (\Omega,t) = \sum_{
	\substack{ m,n \\ 
		m\neq n}} \sum_{\bm{k}}
f_{n\bm{k}}|\alpha_{mn}^{\bm{k}}(t)|^2\delta(\epsilon_{m\bm{k}}-\epsilon_{n\bm{k}}-\Omega) .
\label{eq:spectral}
\end{equation}
This quantity 
contains all electronic excitations between states at different energies in the material. To distinguish the excited electrons, we exclude the projection of wave functions {on} their ground state wave function in the summation. 

The spectral density of the excited state {carries} information such as, at what energies electrons are excited. In Fig.~\ref{fig:spectral}(a) and (b) we present $\eta (\Omega)$  calculated {for the TDDFT and IP approaches,} for the laser with weak and strong intensities, respectively. The spectral distribution shown is computed at $t= 20$ fs. The difference $\Delta \eta$ from these two approaches is shown by blue lines in Fig.~\ref{fig:spectral}(b) and (d). We observe
that electronic excitations happen {dominantly} at the frequencies of the laser field. The imbalance between the peaks at +$\omega_0$ and -$\omega_0$ indicates absorption.

Electronic excitations close to zero frequency {are clearly} visible in Fig.~\ref{fig:spectral}(c), representing {the} electronic Raman transitions {that cause the demagnetization}. The {difference} in $\eta$ {for the TDDFT and IP approaches} shows that electronic Raman transitions are enhanced {by} dynamical modifications in the electronic correlations.  

\section{Comparison of spin and orbital demagnetizations}
\label{appendix:C}
\begin{figure}[t!]
	\centering
     \vspace*{0.3cm}
	\includegraphics[width=\linewidth]{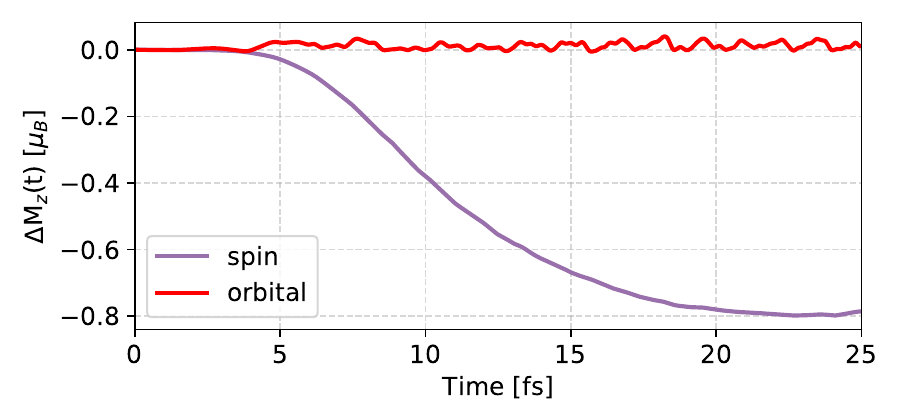}
	\caption{Calculated change in the spin and orbital magnetizations during the ultrafast demagnetization of FePt. 
 The laser parameters used are the same as in Fig.~\ref{fig:fig1}(b).}
	\label{fig:orb}
\end{figure}

Figure \ref{fig:orb} compares the calculated change in spin and orbital angular momentum during ultrafast demagnetization. The laser parameters are the same as in Fig.~\ref{fig:fig1}(b). It is evident that the change in orbital angular momentum is comparatively small. Specifically, in TDDFT the lost angular momentum in the spin part is not 
transferred to the orbital angular momentum of the electrons.

\section{Comparison to demagnetization in bcc Fe}\label{appendix:D}
\begin{figure}[t!]
	\centering
	 \includegraphics[width=\linewidth]{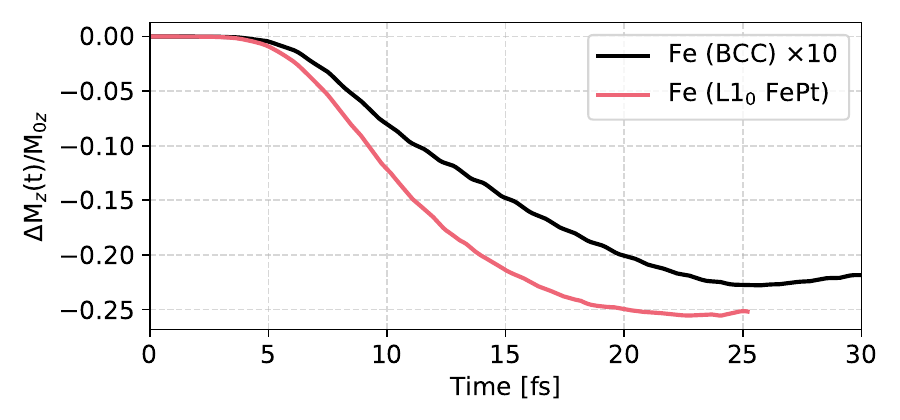}
	\caption{Ultrafast (normalized) demagnetization of Fe in L1$_0$ FePt (pink curve) and of bcc Fe (black curve).  The laser parameters used are the same as in Fig.~\ref{fig:fig1}(b).}
	\label{fig:Fe}
\end{figure}

Figure \ref{fig:Fe} compares the ultrafast demagnetization of Fe in L1$_0$ FePt with that of iron in the  body centered cubic (bcc) unit cell.  The unit cell of iron is modeled with a lattice parameter of 2.48 {\AA}. The bcc Brillouin zone is sampled with a grid of 12 $\times$ 12 $\times$ 12. The demagnetization of Fe in FePt is about ten times larger than that in bcc Fe. The main reason for the stronger demagnetization is the strong spin-orbit coupling of Pt which is favorable for spin-flip Raman-type excitations.

\begin{acknowledgments}
This work has been supported by the Swedish Research Council (VR), the Carl Tryggers Foundation, the German Research Foundation (Deutsche Forschungsgemeinschaft) through CRC/TRR 227 ``Ultrafast Spin Dynamics" (project MF), and the K.\ and A.\ Wallenberg Foundation (Grant No.\ 2022.0079). Part of the calculations were enabled by resources provided by the Swedish National Infrastructure for Computing (SNIC) partially funded by the Swedish Research Council through grant agreement No.\ 2018-05973.
\end{acknowledgments}

%

\end{document}